\documentclass[12pt]{article}
\usepackage{mheckII}

\usepackage{manfnt}
\usepackage{longtable}

\usepackage{blkarray}

\usepackage{tikz} 
\usetikzlibrary{matrix}

\theoremstyle{theorem}

\newtheorem{fact}{Fact}

\theoremstyle{definition}
\newtheorem*{defn}{Definition}
\newtheorem*{rem}{Remark}

\newtheorem{exe}{Example}

\def\Dsl{\,\raise.15ex\hbox{/}\mkern-13.5mu D}
\def\dsl{\,\raise.25ex\hbox{/}\mkern-10.5mu \partial}

\title{$G_2$ Holonomy, Taubes' Construction of Seiberg-Witten Invariants and Superconducting Vortices
}

\authors{Sergio Cecotti\footnote{e-mail: {\tt cecotti@sissa.it}}${}^\S$, Chris Gerig\footnote{e-mail: {\tt cgerig@math.harvard.edu}}${}^{\pmb\star}$, and Cumrun Vafa\footnote{e-mail: {\tt vafa@g.harvard.edu}}${}^\$$\vskip 9pt

\centerline{${}^\S$ SISSA, via Bonomea 265, I-34100 Trieste, ITALY}
\centerline{${}^{\pmb\star}$ Department of Mathematics, Harvard University, Cambridge MA 02138, USA}
\centerline{${}^\S$ Department of Physics Jefferson Physical Laboratory,}
\centerline{Harvard University, Cambridge, MA 02138, USA}}

\abstract{
Using a reformulation of topological ${\cal N}=2$ QFT's in M-theory setup, where QFT is realized via M5 branes wrapping co-associative cycles in a $G_2$ manifold constructed from the space of self-dual 2-forms over a four-fold $X$, we show that superconducting vortices are mapped to M2 branes stretched between M5 branes.  This setup provides a physical explanation of Taubes' construction of the Seiberg-Witten invariants when $X$ is symplectic and the superconducting vortices are realized as pseudo-holomorphic curves.  This setup is general enough to realize topological QFT's arising from ${\cal N}=2$ QFT's from all Gaiotto theories on arbitrary 4-manifolds.}

\begin{document}
\maketitle

\tableofcontents

\newpage 

\section{Introduction}
Topological QFT's introduced by Witten 
\cite{WittenCS,WittenTFT} have been approached from various viewpoints.  A particularly insightful connection has been to realize these within string theory.
In this setup, topological amplitudes can naturally be realized by low energy degrees of freedom living on supersymmetric branes \cite{BSV}. 
A nice set of examples are 3d Chern-Simons theories viewed as theories living on A-branes of topological strings \cite{Witten:1992fb}.  In this setup, one considers the local CY 3-fold string geometry to be $T^*\mspace{-1mu}M^3$ and wraps a D-brane around $M^3$.  Realizing this theory in M-theory \cite{GV,OV}, where M5 branes are wrapped around $M^3\times R^3$, has led to interesting predictions about the integral structure of knot invariants, as well as its extension \cite{GSV} to Khovanov invariants.

Motivated by the connection between superstrings and M-theory, where the strings are mapped to M2 branes, an uplift of topological strings to M-theory, called `topological M-theory' \cite{Dijkgraaf:2004te} was proposed, which replaces CY manifolds with $G_2$ manifolds. 
In this theory one would consider M2 branes wrapping associative 3-cycles instead of holomorphic curves.  Indeed, viewing CY times a circle as a special case of a $G_2$ manifold, the associative cycles are nothing but holomorphic curves times an extra circle. One can also consider the lift of Lagrangian D-branes to topological M-theory. The Lagrangian cycles of topological strings map to co-associative 4-cycles for a $G_2$ manifold. The worldsheet ending on D-branes in topological strings gets mapped to associative subspaces ending on co-associative cycles.  It is natural to ask whether this story has any connections with topological field theory, such as 3d CS theory.  If so it is natural to expect it to be related to a 4d TQFT as 3d Lagrangian subspaces are being replaced by 4d co-associative cycles.

Given the success of 3d TQFT and its relation to topological strings it is natural to ask whether a similar idea would work for 4d TQFT.  In particular it is natural to ask whether the computation of Seiberg-Witten invariants \cite{Witten:1994cg}, which for symplectic manifolds get related to Gromov invariants by Taubes \cite{tau2,tau3}, can be understood from this perspective (see also \cite{TaubesGrSW}).
To obtain a topological theory, as in the 3d case, we need $X$ to be a supersymmetric 4-cycle in a supersymmetric background.  As discussed in \cite{BSV} the natural options are supersymmetric 4-cycles in CY 4-folds, or co-associative cycles in $G_2$ holonomy manifolds, or Cayley 4-cycles in \textsf{spin(7)} manifolds.  However, if we wish to use M5 branes then to get an $\cn=2$ supersymmetric theory on $X$ we need to wrap them in two extra directions, that is, on a Riemann surface $C$.  $C$ must be part of a supersymmetric manifold; for arbitrary $C$ the smallest-dimensional ambient space in which $C$ is calibrated is $T^*\mspace{-1mu}C$.  So the only possibility, given that the dimension of M-theory is 11, is that the 4-cycle $X$ is part of a $G_2$ manifold.  The local structure of the $G_2$ manifold is obtained by considering the space of self-dual 2-forms on $X$ which leads to a 7-fold $W$.  We would then consider M-theory on the 11-dimensional manifold $W\times T^*C$ and wrap M5 branes around $X\times C$.  The low energy, supersymmetric partition function of this theory is naturally captured by ${\cal N}=2$ TQFT of the $4d$ QFT labeled by the curve $C$ on the 4-manifold $X$.  Precisely this geometric realization of 4d TQFT in M-theory, using the $G_2$ structure has already been constructed
and studied in \cite{gukov2}.  Indeed many elements of what we encounter in this paper have been considered there as well.\footnote{A different embedding of 4d TQFT in string theory which has led to insight about their structure was recently considered in \cite{gukov3}.}

The case of the Seiberg-Witten geometry near the monopole point is captured in this setup by the curve $C: x y -a=0$, with $x,y\in \C$, where the monopole point corresponds to $a=0$.
The light monopole, as $a\rightarrow 0$, is realized in M-theory as a M2 brane whose boundary ends on the vanishing cycle as in the setup studied in \cite{KLMVW}.  However, at $a=0$ a new possibility arises:  The curve $C$ splits in two parts and they can be separated.  This corresponds to deforming the $U(1)$ gauge theory by an FI D-term for each harmonic form in $X$.  If the harmonic form has no zeros, as is the case with symplectic $X$, it Higgses\footnote{For the mathematically minded readers, the mechanism of gauge symmetry breaking is nicely explained in \cite{WittenHiggs}.} the $U(1)$.  In particular the co-associative cycle splits into two:  $\{x=0\} \times X \cup \{ y=0\} \times X^4$ separated by the harmonic form in the normal direction to $X$.  The supersymmetric partition function in this case receives contributions only from supersymmetric M2-branes which are in the limit of small separation, when $X$ is symplectic, the same as pseudo-holomorphic curves times an interval along the normal direction as has been shown in \cite{counting}.  Thus the contributions to the partition function of the topological theory become equivalent to studying Gromov-Witten invariants on $X$.  Even if $X$ is not symplectic, this deformation (which is possible only if $b^{+}_2>0$) is still useful, and in this case we will separate the two sides except over the zeros of the harmonic form where the two pieces intersect.  From the physics setup it is clear that we still should be able to compute the partition function in this case, but there would be extra configurations to take into account.  This is in accord with recent results in \cite{gerig1,gerig2} which show that one needs to include pseudo-holomorphic curves which end on the zeros.  This is natural because this still gives rise to the M2 branes ending on the M5 branes.  Even if $b^{2+}=0$ and we could not deform the curves, this setup is still valid, but does not lead to any simple way to compute it as the light modes are no longer localized to pseudo-holomorphic curves in $X$.

This setup naturally extends to Gaiotto $\cn=2$ theories where we wrap $N$ M5 branes over the Seiberg-Witten (SW) curve.  To apply this setup we need a family of non-compact SW curves $C_u$, parametrized by the $\cn=2$ Coulomb branch $\cu$, such that there are points $u_\star\in\cu$ where $C_{u_\star}$ degenerates into nodal genus 0 curves touching at points.  Along this degenerating locus the topological amplitudes typically diverge. However, in such cases the local theory would have $U(1)^k$ global symmetry where $k$ is the number of double points in $C_{u_\star}$.  If in addition we gauge this symmetry, we can introduce FI D-terms in the corresponding $U(1)$'s which removes the singularities, which would correspond to separating $C$ into disconnected genus 0 pieces, and the above Seiberg-Witten geometry applies locally to all such points and leads to computation of the corresponding topological amplitudes.

The organization of this paper is as follows:  In section 2 we introduce the geometric setup.  In section 3 we explain the physical interpretation of the setup and the deformation.
Finally, in section 4 we end with some conclusions.

\section{The geometric setup}\label{setup}

In this note we consider M-theory on a Euclidean 11-fold of the form
\be
M_{11}= W\times H,
\ee
where $W$ is a 7-fold of $G_2$ holonomy with parallel 3-form $\Phi_3$ \cite{g21,g22}, and $H$ a hyperK\"ahler 4-fold; neither space is supposed to be compact or complete.  This geometric compactification of M-theory has been considered in \cite{gukov2}.
 This geometry preserves two supersymmetries. Let $L_0$ be a calibrated submanifold of the form
\be
L_0\cong X\times C\subset W\times H
\ee
where $X\subset W$ is a compact co-associative submanifold (i.e.\! $\Phi_3|_X=0$) \cite{g22} and $C\subset H$ a special Lagrangian submanifold which is a holomorphic curve in complex structure $I$.   Let $U\subset X$ be a coordinate patch\footnote{\ To the best of our knowledge, it is not known if there exist \emph{global} obstructions to the isometric embedding of an arbitrary orientable Riemannian 4-manifold $X$ as a co-associative submanifold of some (non-complete) $G_2$-manifold; if present, they are expected to be quite mild \cite{bryant}.} in any real-analytic Riemannian 4-fold $X$; we can always find a (non-complete) $G_2$ manifold $W_U$ with an anti-$G_2$ involution $r$
(i.e.\! $r^*\Phi_3=-\Phi_3$) so that $U$ embeds isometrically in $W_U$
as the co-associative submanifold $\mathsf{Fix}(r)$ of the fixed points of $r$, in fact we may even choose $W_U$ so that the embedding is totally geodesic \cite{bryant}.
We are interested in the local physics near $L_0$, and we may replace $W\times H$ by a tubular neighborhood of $L_0$ which is isomorphic to 
the total space of the bundle \cite{lean}
\be
W\mspace{-2mu}X\times T^*\mspace{-1mu}C\to L_0
\ee
where $W\mspace{-2mu}X\to X$ is the vector bundle
of self-dual 2-forms and $T^*\mspace{-1mu}C\to C$ the canonical bundle.  We identify $X$ with the zero section $s_0\colon X\to W\mspace{-2mu}X$, and write $g$ for the genus of $C$. The $G_2$-structure along $X\subset W\mspace{-2mu}X$  is modeled on the 3-form
\be
\Phi_3=\upsilon- \eta^a_{\ \mu\nu}\, dw_a\wedge dx^\mu\wedge dx^\nu
\ee
where $w_a$ ($a=1,2,3$) are coordinates along the fiber, $x^\mu$ local coordinates in  $X$, $\upsilon$ the volume form of the fiber, and $\eta^a_{\ \mu\nu}$ the 't Hooft tensor \cite{tHooft:1976snw}.
The map $\xi\mapsto s_0^*(i_\xi\Phi_3)$ identifies isomorphically the tangent space to the fiber of $W\mspace{-2mu}X$ with the space of self-dual 2-forms on $X$. The form $\Phi_3$ together with the $G_2$ metric $G$ define
a vector cross product $\times$ on the tangent bundle $TW\mspace{-2mu}X$ preserved by parallel transport
\be\label{vcp}
\times\colon TW\mspace{-2mu}X \wedge TW\mspace{-2mu}X\to TW\mspace{-2mu}X,\qquad 
G(u\times v,w)=\Phi_3(u,v,w). 
\ee

We start by wrapping a M5-brane on  the
6-dimensional space
\be\label{jjjx}
L=X\times \big\{y^N-y^{N-2}\phi_2+y^{N-3}\phi_3+\cdots\pm\phi_N=0\big\}\subset W\mspace{-2mu}X\times T^*\mspace{-1mu}C,\qquad N\geq 2
\ee
where $y$ is a fiber coordinate for $T^*\mspace{-1mu}C$ and $\phi_k$ a meromorphic $k$-differential on $C$.

If $X$ is flat and very large, the 4d IR world-volume theory on $X$ is just the class-$\cs$ 4d $\cn=2$  Gaiotto theory \cite{Gaiotto:2009hg,Gaiotto:2009we} defined by the data $(C,\{\phi_k\})$ quantized in the Euclidean 4-manifold $X$ (plus a decoupled free theory for the center of mass d.o.f.). In the flat case the 4d theory preserves 8 supercharges in the representation $(\boldsymbol{2},\boldsymbol{1},\boldsymbol{2})_{+1}\oplus 
(\boldsymbol{1},\boldsymbol{2},\boldsymbol{2})_{-1}$ of the $\text{(Lorentz)}\times\text{(R-symmetry)}$ group
 \be
 SU(2)_+\times SU(2)_-\times SU(2)_R\times U(1)_r.
 \ee
 The symmetry $SU(2)_R$ is geometrically identified with the rotations of the $\R^3$ fiber of the bundle $W\mspace{-2mu}X\to X$. The $G_2$-structure identifies the fiber $W\mspace{-2mu}X_x$ with the vector space of self-dual 2-forms at the point $x\in X$, and hence $SU(2)_R$ with the self-dual factor $SU(2)_+$ in the 4d Euclidean Lorentz group. The supercharge $\cq$ invariant under 
$SU(2)_\text{diag}\subset SU(2)_+\times SU(2)_R$ is the topological supersymmetry of the $\cn=2$ theory topologically twisted \emph{\'a la} Witten \cite{WittenTFT} (for a nice survey see the book \cite{Marino}, and for the geometric setup relevant for our discussion see \cite{gukov}). 
The supersymmetry $\cq$ remains unbroken even when $X$ is curved. More generally, the topological supersymmetry $\cq$ is preserved by all deformations of the 4-manifold $X$ inside  $W\mspace{-2mu}X$ as long as the deformed space $X_\text{def}$ is a co-associative submanifold of $W\mspace{-2mu}X$ since the $G_2$-structure identifies the $SU(2)_+$ and $SU(2)_R$ connections on $X_\text{def}$ and one covariantly constant susy parameter $\epsilon$ is still present. 

$\cq$ is nilpotent, $\cq^2=0$, and the topological states/operators are $\cq$-cohomology classes. Each observable $\co$ has a $k$-form version $\co^{(k)}$ for all $k$ so that
their integrals on $k$-cycles $\co(\Gamma_k)\equiv \int_{\Gamma_k}\co^{(k)}$ are $\cq$-closed \cite{WittenTFT,Marino}. The quantities of main interest are the topological correlation functions
\be\label{hhasqw}
\Big\langle \co_{i_1}(\Gamma_{k_1})\, \co_{i_2}(\Gamma_{k_2})\cdots \co_{i_\ell}(\Gamma_{k_\ell})\Big\rangle_{\!X}
\ee
which are topological invariants of the smooth 4-manifold $X$.
\medskip

Under certain geometric conditions (to be specified in a moment) the M5 brane configuration \eqref{jjjx} admits an interesting deformation which preserves the topological supersymmetry $\cq$. The goal of the present note is to give a novel interpretation of this deformation and study some of its implications. We shall proceed by steps.

\subparagraph{The deformation of $X$ in $W\mspace{-2mu}X$.} Let us deform $X$ to a nearby 4-fold
 $X_\text{def}\subset W\mspace{-2mu}X$
 specified locally by the equation 
 \be\label{kkkz10i}
 w_a=\varepsilon\, \phi_a(x)+O(\varepsilon^2).
 \ee 
 One has
 \be
 \Phi_3\big|_{X_{\text{def}}}=\varepsilon\, d\big(\eta^a_{\ \mu\nu}\, \phi_a(x)\,dx^\mu\wedge dx^\nu\big)+O(\varepsilon^2)=0
 \ee
 so, to the first order in $\varepsilon$, a deformed co-associative submanifold $X_\text{def}$ is just the graph $X_\omega$ of a closed self-dual, hence harmonic, 2-form 
 \be
 \omega=\eta^a_{\ \mu\nu}\, \phi_a(x)\,dx^\mu\wedge dx^\nu.
 \ee
One shows\footnote{\ See e.g.\!  \S4 of \cite{lean} or \S12.3.1 of \cite{g22}.} that this deformation is not obstructed to higher order, so it make sense to speak of the deformation 
\be\label{defff}
X\leadsto X_\omega
\ee 
by a finite\footnote{\ At least as long as $\|\omega\|^2$ is not too large.}
self-dual harmonic 2-form $\omega$: the 4-fold $X_\omega\subset W\mspace{-2mu}X$ is compact and co-associative.
The deformation space of $X$ is smooth of real dimension \be
b_2^+(X)=\dim_\R H^{2}(X,\R)^+.
\ee  
To have non-trivial deformations, in this paper we shall always  assume  
$b_2^+(X)\geq1$. To get a simpler theory it is sometimes convenient to assume the stronger condition $b_2^+(X)>1$.\footnote{ Since the theory is topological, in fact, partially topological since the topological correlation functions depend on non-normalizable complex deformations of non-compact $C$ which corresponds to masses, we may as well consider the opposite limit, namely $X$ small and $C$ very large. From this alternative  point of  view we get the 2d TFT on $C$ obtained by twisting the 2d $(2,0)$ model associated to the 4-fold $X$, see 
refs.\!\cite{gukov,gukov2}. However the deformation we are interested in seems more naturally described from the perspective of TFT on the space-time $X$.}
\medskip

\subparagraph{The factorization locus $\cu_\circ\subset\cu$.}
The coefficients $\phi_k$ of the Seiberg-Witten (SW) curve for the underlying 4d $\cn=2$ model
\eqref{jjjx}
\be\label{swccc}
y^N-y^{N-2}\phi_2+y^{N-3}\phi_3+\cdots\pm\phi_N=0
\ee
depends on fixed parameters, such as the masses, as well as 
on the point $u$ in the Coulomb branch $\cu$ over which one has to integrate because the Euclidean space-time $X$ is compact.
Contrary to the usual treatment \cite{integration,Marino}, 
we require the fixed parameters to have \emph{non-generic} 
values such that there is a non-empty sub-locus $\cu_\circ\subset \cu$ where the SW curve is maximally reducible into $N$ distinct components, i.e.\! it splits into linear factors
\be\label{kasqwz}
y^N-y^{N-2}\phi_2\big|_{\cu_\circ}+y^{N-3}\phi_3\big|_{\cu_\circ}+\cdots\pm\phi_N\big|_{\cu_\circ}=\prod_{\ell=1}^N(y-\lambda_\ell),\qquad \sum_{\ell=1}^N\lambda_\ell=0,
\ee 
where $\lambda_\ell$ are meromorphic differentials on $C$
($\lambda_\ell\not\equiv\lambda_{\ell^\prime}$ for $\ell^\prime\neq\ell$).

Formulae simplify in the $N=2$ case where eqn.\eqref{kasqwz} reduces to
\be\label{sqqare}
\phi_2\big|_{\cu_\circ}=\lambda^2
\ee 
for some meromorphic differential $\lambda$, and 
$\cu_\circ\neq\varnothing$ iff $(C,\phi_2)$ satisfies two conditions:
\begin{itemize}
\item[C1.] $\phi_2$ has poles of \emph{even} order $2n_i$ at finitely many punctures $z_i\in C$ ($i=1,\dots,p$), i.e. 
\be\label{res1}
\sqrt{\phi_2(u;z)}=\pm\sum_{s=1}^{n_i} \frac{\Lambda_{i,s}}{(z-z_i)^s}\,dz+\text{regular as }z\to z_i,\qquad \Lambda_{i,n_i}\neq0.
\ee
The positive integers $n_i$ are restricted by the condition that the dimension $k$ of the Coulomb branch $\cu$ of the 4d $\cn=2$ theory is non-negative
\be
k\equiv\dim_\C\cu= 3(g-1)+\sum_{i=1}^pn_i\geq0.
\ee 
$\lambda$ has poles of order $n_i$
at $z_i$ whose principal parts are as in  eqn.\eqref{res1};
\item[C2.] 
for some choice of $\epsilon_i=\pm1$, the mass parameters $m_i\equiv \Lambda_{i,1}$
satisfy
\be\label{res2}
\sum_{i=1}^p\epsilon_i\, m_i=0.
\ee
Eqn.\eqref{res2} reflects the fact that the total residue of the meromorphic 1-form $\lambda$ vanishes.
\end{itemize} 
In the $N=2$ case, when $\phi_2$ is holomorphic the 6-fold $L$ in \eqref{jjjx} has the form 
$X\times \text{(compact)}$.
We are mainly interested in the opposite situation where the SW curve $\{y^2=\phi_2\}$ is non-compact: this requires at least one puncture to be present.
$\cu_\circ=\cup_i\,\cu_i$ decomposes in finitely many  irreducible components such that
\be
\cu_i\cong \C^g\ \text{as complex manifolds}.
\ee 
In particular for $g=0$ the locus $\cu_\circ\subset \cu$ consists of finitely many points.
\medskip  

The class-$\cs[A_1]$ QFT specified by the datum $(C,\phi_2)$ has a Lagrangian formulation when $C=\bP^1$ and $\phi_2$ has a single pole with $n_1=3$, or for $C$ arbitrary and $n_i\in\{1,2\}$ \cite{Gaiotto:2009we,Cecotti:2011rv}.
In the second case the flavor symmetry is at least\footnote{\ There are 5 special cases where the symmetry enhances to a larger group of the same rank \cite{Cecotti:2011rv}.}
$
SU(2)^p
$. 
The masses $m_i$ take value in the Cartan subalgebra $\mathfrak{h}$ of the flavor symmetry.

\begin{exe} For instance, if the underlying class-$\cs[A_1]$ model is SQCD with $N_f=2$ (which corresponds to $(n_1,n_2)=(2,2)$)
with quark masses $m_1=\pm m_2=m$, we have
\be
\phi_2(u;z)=\left(\frac{\Lambda^2}{z^4}+\frac{2\Lambda m}{z^3}+\frac{4u}{z^2}\pm\frac{2\Lambda m}{z}+\Lambda^2\right)\!dz^2,
\ee
and $\cu_\circ$ consists of the single point $u_\circ=(m^2\pm 2\Lambda^2)/4$. At $u_\circ$
eqn.\eqref{sqqare} holds with 
\be
\lambda=\left(\frac{\Lambda}{z^2}+\frac{m}{z} \pm \Lambda\right)\!dz.
\ee
\end{exe}

\subparagraph{Co-associative deformations of the M5 branes.}
We return to the general case of a SW curve satisfying the maximal factorization property \eqref{kasqwz}
(but otherwise generic). On the locus $\cu_\circ\subset\cu$
the M5 support $L$, eqn.\eqref{jjjx}, becomes reducible
\be
L=\bigcup_{\ell=1}^NL_\ell,\qquad L_\ell\equiv X\times \big\{y=\lambda_\ell\big\}\subset W\mspace{-2mu}X\times T^*\mspace{-1mu}C,
\ee  
and we can separate the various irreducible components in the $W\mspace{-2mu}X$ direction
\be
L\leadsto \bigcup_{\ell=1}^N L_{\ell,\omega_\ell},\qquad
L_{\ell,\omega_\ell}\equiv X_{\omega_\ell}\times \big\{y=\lambda_\ell\big\}\subset W\mspace{-2mu}X\times T^*\mspace{-1mu}C,
\ee
where $\omega_\ell\in \Omega_2^+(X)$ are distinct self-dual harmonic forms.
This is the deformed M5 brane configuration we are interested in. By construction it still preserves the 
topological supersymmetry $\cq$.
\medskip

For ease of presentation, from now on we focus on the $N=2$ case, the extension to general $N$ being clear. The underlying 4d $\cn=2$ QFT is then of class-$\cs[A_1]$. For $N=2$ the support of the M5 branes is simply
\be\label{brannessup}
L=L_\omega\cup L_-,\quad\text{where}\quad L_\omega= X_\omega\times \big\{y=\lambda\big\},\quad
L_-=X\times \big\{y=-\lambda\big\}.
\ee 
The intersection $\{y=\lambda\}\cap\{y=-\lambda\}$ generically consists of 
\be
h=2(g-1)+\sum_i n_i
\ee 
distinct points ($\equiv$ double zeros of $\phi_2$).
\medskip

In a general class-$\cs[A_1]$ QFT, when we approach a point $u\in\cu$ where $\phi_2(u)$ has a zero of order 2 (which may be thought of as the result of the collision of two simple zeros), a hypermultiplet becomes massless and we need to insert it in the IR description. Approaching a zero of higher order
 the massless hypermultiplet gets replaced by a strongly interacting Argyres-Douglas (AD) SCFT \cite{AD} which also  becomes part of the IR physics.

In our setup, as we approach the special locus $\cu_\circ\subset\cu$, all zeros of $\phi_2$ get of even order. Approaching a generic point in $\cu_\circ$, $h$ mutually-local hypermultiplets get massless. In codimension 1 in $\cu_\circ$, interacting AD systems also enter in the IR description. The emergence of AD SCFT's is then generic for $g\geq1$. For $g=0$ with general masses satisfying \eqref{res2} no AD system appears anywhere in the Coulomb branch $\cu$.
To further simplify the discussion, we focus on $g=0$ with arbitrarily many punctures satisfying C1, C2.
Then $\cu_\circ\subset \cu$ is a finite collection of points.
As we approach a factorization point $u_\circ\in \cu_\circ$,
$h\equiv \sum_in_i-2$ mutually-local hypers get light;
since we have only $k\equiv \sum_i n_i-3$ photons,
a Higgs branch of quaternionic dimension 1 opens up at each $u_\circ\in\cu_\circ$.

\begin{exe}\label{nnnzxa1} The simplest possible instance is $C=\bP^1$ with a single pole with $n_1=3$, that is,
$\phi_2= z^2 dz^2$. In this case $h=1$ and $k=0$, so the underlying $\cn=2$ QFT is just a free massless hypermultiplet. The two M5 branes have support
\be\label{nnn67we}
X_\omega\times \{x-y=0\}\quad
\text{and}\quad X\times \{x+y=0\}.\ee
\end{exe}

\subsection{Some useful geometric facts}

\subsubsection{$\omega$ symplectic}\label{se:sym} 

In the special case that the self-dual harmonic form $\omega$ is actually a symplectic form (i.e.\! it vanishes nowhere)
the supports of the two M5 branes \eqref{brannessup} are completely separated
\be\label{jjaqw}
L_\omega\cap L_-=\varnothing.
\ee 

We write $\omega= t\, \Omega$, $t\in \R$, where the self-dual symplectic form $\Omega$ is normalized so that $\|\Omega\|^2=2$. There exists a compatible almost complex structure
$J\colon TX\to TX$, $J^2=-1$, such that
the Riemannian metric $G$ has the form \cite{salamon}
\be\label{jdefff}
G(v,w)=\Omega(v, Jw).
\ee
When $J$ is integrable the metric $G$ is K\"ahler with K\"ahler form $\Omega$. 
In general, $J$ decomposes the complexified differential forms into $(p,q)$-type
\be
\wedge^k T^*\mspace{-2mu}X\otimes\C=\bigoplus_{p+q=k} T^{(p,q)},\quad T^{(p,q)}=\wedge^p T^{(1,0)}\otimes \wedge^q T^{(0,1)},\quad T^*\mspace{-2mu}X\otimes \C=T^{(1,0)}\oplus T^{(0,1)}.
\ee
The canonical line bundle is $K=T^{(2,0)}$ and we write $c$ for its Chern class $c_1(K)$.

A compact 2-dimensional submanifold $\Sigma\subset X$ is called a \textit{pseudo-holomorphic curve} iff $J$ preserves $T\Sigma$. In this case $J$ induces on $\Sigma$ the structure of a complex curve, the inclusion $i\colon \Sigma\to X$ is a pseudo-holomorphic map in the sense of Gromov \cite{gro}, 
and $\Omega|_{\Sigma}$ is the induced volume form on $\Sigma$. 
We write $e=e(\Sigma)$ for the 2-form Poincar\'e dual to the fundamental class of $\Sigma$; the volume of the pseudo-holomorphic curve $\Sigma$ is
\be\label{vooolw}
\mathrm{vol}(\Sigma)=\int_X \Omega\wedge e(\Sigma).
\ee
If $\Sigma$ is connected, its genus is
\be
g(\Sigma)= 1+\frac{1}{2}\big(e\cdot e+c\cdot e\big),
\ee
while the formal dimension of the deformation space of $\Sigma$ in $X$ is \cite{tau1,tau2}
\be
2d= e\cdot e-c\cdot e.
\ee

\subsubsection{$\omega$ near-symplectic}

For a generic metric on a compact 4-fold $X$ with $b_2^+(X)\geq1$, the zero set of a self-dual harmonic 2-form $\omega$  is a finite collection of non-intersecting codimension-3 circles $\amalg_\alpha S^1_\alpha\subset X$ \cite{circles,thesis},
so that the intersection between the two M5's takes the form 
\be\label{ccircles}
L_\omega\cap L_-=\coprod_{\alpha,a} S^1_\alpha\times\{q_a\}\subset X\times C,
\ee 
where $\{q_a\}\subset C$ are the zeros of $\lambda$. We shall refer to this situation as the \emph{near-symplectic} case. One shows that for a generic metric one can choose the self-dual harmonic form $\omega$ so that it has a single circle of zeros \cite{taubeszero}.
To fix the ideas, we assume this choice. 

We cut out a tubular neighborhood $T_\epsilon$ of the zero set $S^1\subset X$ of radius $\epsilon$. We remain with a symplectic 4-manifold $\mathring{X}_\epsilon=X\setminus T_\epsilon$ with boundary $\partial\mathring{X}_\epsilon\cong S^1\times S^2_\epsilon$. The boundary $\partial\mathring{X}_\epsilon$ inherits a contact structure from the symplectic structure in the bulk \cite{taubes4}. The symplectic geometry of the manifold $\mathring{X}_\epsilon$ with boundary $\partial\mathring{X}_\epsilon$ contains a new interesting class of pseudo-holomorphic curves $\Sigma$,
namely the ones with boundaries on 
$\partial\mathring{X}_\epsilon$ which have finite area and satisfy some good boundary conditions \cite{taubes5,gerig2}. Each component of the boundary $\partial\Sigma\subset\partial\mathring{X}_\epsilon$ is a (multiple cover of a) closed curve $\gamma$ in the contact 3-fold $\partial\mathring{X}_\epsilon$:  the appropriate boundary condition is that, as $\epsilon\to0$, the curve $\gamma$ approaches an orbit of the Reeb vector field for the induced contact structure \cite{taubes5,gerig2}.
One shows that if the Seiberg-Witten invariants of the 4-manifold $X$ are not zero, there must be such finite-area
pseudo-holomorphic curves with Reeb orbit boundaries. In fact, one may recover the Seiberg-Witten invariants by a suitable count of such curves \cite{gerig2}.         

\section{Physical interpretation of the deformation} 

\subsection{Generalities}\label{phint}

When the differential $\lambda$ is holomorphic, the deformation $L\leadsto L_\omega$ is normalizable and the deformation parameter  
$\omega$ is a dynamical field from the viewpoint of the 4d QFT on $X$. If $\lambda$ has non-trivial poles (as is automatically the case for $g=0$)
the deformation is non-compact and $\omega$ becomes a frozen parameter from the 4d perspective. Formally we may still consider $\omega$ as a component of a (non-dynamical) background supermultiplet in the same $\cn=2$ susy   representation as its compact-case counterpart. Since $\omega\neq0$ does not break the topological supersymmetry, $\omega$ should be the v.e.v. of the lowest component in its supermultiplet. In geometric engineering of $\cn=2$ theories, the R-symmetry is identified with the group of automorphisms of the normal bundle to the world-brane; it follows that $\omega$ transforms as a triplet under $SU(2)_R$ (identified with $SU(2)_+$ by the topological twist).  The obvious $\cn=2$ supermultiplet whose first component is a $SU(2)_R$ triplet is the linear one, i.e.\! the supermultiplet containing a conserved flavor current $J_f^\mu$. The first component of the linear  supermultiplet is the triplet of hyperK\"ahler moment maps of the corresponding flavor symmetry. The linear supermultiplet contains a 2-form gauge field $B$, related to the flavor current by $J_f=\ast dB$. The 2-form $B$ may be identified with a non-normalizable mode of the 2-form living on the M5 world-volume. 

An $\cn=2$ susy-preserving coupling which may be interpreted as a background linear multiplet is nothing else than an $\cn=2$ Fayet-Iliopoulos (FI) term
for an abelian vector-multiplet which may be made of fundamental fields, composite operators, or non-dynamical degrees of freedom.

We are thus led to consider FI terms of abelian gauge theories.  The FI deformation of topological theory under consideration has also been considered in \cite{gukov2}.
 \subsection{Topological FI terms}\label{fififi}

We recall that, after the topological twist, the components of a $\cn=2$ vector-multiplet are:
a gauge vector $A_\mu$, a complex scalar $\phi$,
an auxiliary field $D$ which is a real self-dual 2-form,
a one-form fermion $\psi$, a self-dual 2-form fermion $\chi$, and a scalar fermion $\eta$ (all fields being in the adjoint of the gauge group). We write $\delta$ for the action of the topological supersymmetry. In particular we have\footnote{\ These formulae hold modulo gauge transformations \cite{Marino}.}
\be\label{deeelt}
\delta\phi=0,\qquad \delta\chi=D-iF^+,
\ee
where $F^+$ stands for the self-dual projection of the field strength $F=dA+A^2$.
\medskip

Let $S$ be the action of a topologically twisted 4d $\cn=2$ theory which contains an abelian vector-multiplet $(\phi, \psi, \chi, \eta, D, A_\mu)$. We may add to $S$ a $\delta$-exact term of the form
\be\label{oop0}
S\to S(\omega)\equiv S+\delta\!\!\int_X\omega \wedge \chi,
\ee 
where $\omega$ is a closed self-dual 2-form.
The modification \eqref{oop0} does not change the topological correlations \eqref{hhasqw} which then are $\omega$-independent. We call the new term in the \textsc{rhs} of  \eqref{oop0} a topological FI coupling.

The topological FI term may be generalized to the non-abelian case 
\be\label{kiqwaer}
S\to S+\delta\!\!\int_X\omega \wedge \mathrm{tr}\big(P^\prime(\phi)\chi\big)
\ee
where $\phi$ is the scalar of a non-abelian vector multiplet, $\chi$ its self-dual 2-form fermion and 
$\mathrm{tr}\,P(\phi)$ stands for any ad-invariant symmetric polynomial. 

Using eqns.\eqref{deeelt}, eqn.\eqref{oop0} becomes
\be\label{oop1}
\begin{aligned}
S\to &S+\int_X \omega \wedge D-i\!\int_X\omega\wedge F=\\
&= S+\int_X \omega \wedge D+2\pi\! \int_X\omega\wedge c_1(\cl),
\end{aligned}
\ee
where $\cl$ is the line bundle associated to the abelian gauge field and we used 
\be
\int_X \omega\wedge F^-=0
\ee 
since $\omega$ is self-dual.
Then, up to the topological term $2\pi \int_X \omega\wedge c_1(\cl)$, the  $\cq$-exact deformation \eqref{oop0} just adds  to the action the FI term
\be
\int_X \omega\; D.
\ee
Therefore the topologically trivial modification \eqref{oop0} has two effects:
\begin{itemize}
\item[a)] it multiplies the topological path integral in each topological sector by the constant
\be
e^{-2\pi [\omega]\cdot c_1(\cl)},
\ee
\item[b)] it modifies the equation of motions of the auxiliary field $D$ with the effect of shifting its  on-shell value: $D_\text{on-sh}\to D_\text{on-sh}-e^2\omega$ where $e$ is the abelian gauge coupling.
\end{itemize}
The statement that the combined effect of a) and b) is to leave the smooth invariants \eqref{hhasqw} unchanged is equivalent to the well-established validity of the usual deformation \cite{Witten:1994cg}\!\!\cite{tau1,tau2} used to simplify the computation of the Seiberg-Witten invariants \cite{Witten:1994cg}\!\!\cite{Marino}
when $b^+_2(X)>1$. (For $b^+_2(X)=1$ the situation is a bit subtler, and some more care is needed \cite{tau1,tau2}). 
\medskip

\noindent In the non-abelian case one may write 
the last term in \eqref{kiqwaer} as the 
topological observable\footnote{\ Here $[\omega]$ stands for the 2-cycle Poincar\'e dual to $\omega$.}
\be
\int_{[\omega]} \mathrm{tr}\,P(\phi)^{(2)},
\ee
 plus a bilinear in the one-form fermion $\psi$ of the vector-multiplet.
 \medskip

 The discussion in \S\ref{phint} suggests that the deformation $L\leadsto L_\omega$ has the effect of modifying the topologically twisted IR effective theory by adding a FI term of the general form
 \be\label{kazqwcx}
 S\to S-\delta\!\!\int_X \omega\wedge \sum_a \kappa_a\, \chi^a+2\pi \sum_a \kappa_a \int_X\omega\wedge c_1(\cl^a), 
 \ee 
 where the sum is over all the light photons and $\kappa_a$ are numerical coefficients which depend on the SW curve and the point $u_\circ\in \cu$. The new action \eqref{kazqwcx} is still topologically invariant.
 We shall make precise applications of this idea in the following subsections to the theories under consideration.
 \medskip

\subsection{M2 branes wrapped on associative cycles}\label{m2wap}

To compute topological correlation functions from our geometric setup we have to sum over all $\cq$-invariant configurations describing finite-action instantons of our system of M5 branes wrapped on $L_\omega\cup L_-$. In the $N=2$ case these instantons are finite-volume BPS M2 branes   
suspended between the two M5 supported on $L_\omega$ and $L_-$.
Such a M2 brane does not break the topological supersymmetry iff each connected component $M\subset W\mspace{-2mu}X\times T^*\mspace{-1mu}C$ of
its support is calibrated. In particular, the projection of $M$
on the first factor space, $W\mspace{-2mu}X$, should be either a point or
a connected associative 3-manifold $A$. Saying that $A$ is associative is equivalent to saying that its tangent space $TA$ is closed under the vector cross-product $\times$
\eqref{vcp} or, equivalently, that is calibrated by $\Phi_3$ i.e.
\be
\Phi_3\big|_A=\upsilon_A\equiv \text{(induced volume form)}.
\ee
We distinguish two cases.

\subsubsection{$\omega$ symplectic}  For $\omega$ symplectic $X_\omega\cap X=\varnothing$,
so the projection of $M$ on $W\mspace{-2mu}X$ cannot be a point, hence it must be an associative 3-fold $A\subset W\mspace{-2mu}X$. Then the projection of each connected component of the M2 brane on $T^*\mspace{-1mu}C$ is a point and
\be\label{supM}
M= A\times \{q\}, \qquad q\in \{y=\lambda\}\cap\{y=-\lambda\}\subset T^*\mspace{-1mu}C.
\ee
It follows that the projection of $M$ on the second space $T^*\mspace{-1mu}C$ must be a zero of the differential $\lambda$. These zeros are in one-to-one correspondence with the hypers which get light as $u\to u_\circ\in \cu_\circ$, so to each connected BPS M2 there is associated a particular massless hyper.
\medskip

As  before, for $\omega$ symplectic we write $\omega=t\Omega$ with $\|\Omega\|^2=2$.
We claim that for small $t$ the boundary of $A$ in $X$ (as well as in $X_{t\Omega}$) is  a pseudo-holomorphic curve $\Sigma$ with respect to the almost complex structure $J$ defined by the self-dual symplectic form $\Omega$, cfr.\! eqn.\eqref{jdefff}. 
Indeed, in a neighborhood of $X$ the associative 3-fold has the form 
\be
A=\Big\{\big(x+O(s), w_a=s\,\phi_a(x)+O(s^2)\big),\quad x\in\Sigma,\  s\geq 0\Big\}\subset W\mspace{-2mu}X
\ee
where $\phi_a(x)$ is as in eqn.\eqref{kkkz10i}.
The vertical subbundle $V_A\subset TA$ is spanned by $\partial_s$ and  
\be
G(\partial_s \times u, v)=\Phi_3(\partial_s, u, v)=\Omega(u, v)=G(J u, v)\qquad
u,v\in TX,
\ee
that is, $J u= \partial_s \times u$ so that $TA$ closed under $\times$ implies that $T\Sigma\simeq TA/V_A$ is closed under $J$, i.e.\! the boundary $\Sigma=\partial A\cap X$ is pseudo-holomorphic. Vice-versa, if we have a pseudo-holomorphic curve $\Sigma\subset X$ we may construct an associative 3-fold $A\subset W\mspace{-2mu}X$ such that $\partial A\cap X=\Sigma$. We conclude that
associative 3-folds suspended between $X$ and $X_{t\Omega}$ and pseudo-holomorphic curves in $X$ are in one-to-one correspondence
(for small $t$ and $X$ symplectic), and counting associative 3-folds $A$ with boundaries on $X$ and $X_\omega$ in a given topological class is equivalent to counting
pesudo-holomorphic curves with given homology class $e(\Sigma)$ (see \cite{counting} for a precise mathematical treatment).
\medskip

As a check, let us compute the volume of the associative submanifold $A$ (to the first order in $t$), cfr.\! eqn.\eqref{vooolw}
\be
\mathrm{vol}(A) =t\cdot \mathrm{vol}(\Sigma)=
t\!\int_X\Omega\wedge e(\Sigma)=\int_X\omega\wedge e(\Sigma).
\ee

Since the Euclidean M2 branes wrapped on an associative manifold $A$ with boundaries in the 
co-associative spaces $X_\omega$, $X$ are BPS, we expect that they give a contribution to the topological action of the form
\be\label{qahhawr}
T\,\mathrm{vol}(A)+\delta\text{-exact}\equiv
T\!\int_X\omega\wedge e(\Sigma)+\delta\text{-exact},
\ee
where $T$ is the M2 tension in the appropriate units. Eqn.\eqref{qahhawr} matches with
the expression we found for the topological FI terms \eqref{kazqwcx} (say with one vector-multiplet) provided the following identifications hold
\be\label{kasqwer}
e(\Sigma)=c_1(\cl),\qquad\quad
T=2\pi \kappa.
\ee
The second condition may be taken to be the definition of $\kappa$. 
To understand the validity of the first identification we will first consider the situation for a $U(1)$ $\cn=2$ theory with a massless charged field, as in the monopole point of the Seiberg-Witten geometry, and then explain how the generalization works for the case in consideration\footnote{For a review of work by Taubes
\cite{tau1,tau2,tau3} on the Seiberg-Witten monopole equations \cite{Witten:1994cg} and their relation with the Gromov invariants \cite{gro} when $X$ is symplectic and $b_2^+(X)>1$, see the appendix.}.

\subsubsection{The $U(1)$ monopole point and its deformation}

In this section let us focus on the original Seiberg-Witten monopole point (example \ref{nnnzxa1} of \S2).  The geometry in this case is represented by $C: \ \ xy =0$.  The $U(1)$ gauge field on the M5 brane arises from the $B$-field on the M5 brane as follows:  Consider a generic Coulomb branch point deformation, given by $xy=\mu$.  In this case we have a non-trivial 1-cycle on $C$ and a dual one form $\eta$.  The $U(1)$ gauge field $A$ on $X$ arises from the M5 brane by decomposing it in the direction of this 1-form: $B=A(x)\wedge \eta$.
Note that the mass of the charged field is $\mu$ which goes to zero as $\mu\rightarrow 0$.  If we turn on the FI term for the $U(1)$ the Coulomb branch is automatically pushed to $0$ to allow the condensation of a v.e.v. for the massless fields to preserve supersymmetry.  This corresponds to $\mu =0$ and then pulling the $xy=0$ curve to two disconnected curves in the full $G_2$ geometry given by $x=0$ and $y=0$ with different transverse positions.  Now there is no compact cycle on $C$ and this corresponds to Higgsing the $U(1)$ via the FI term.  

Let us see how the situation changes in presence of a M2 instanton with world-volume  $\Sigma \times I$ where $I$ is the unit interval whose two ends are, respectively, at the point $x=0$
on the curve $\{y=0\}$ and at the point $y=0$ on $\{x=0\}$.
From the point of view of the world-volume theory on each M5 brane, this instanton is a topological defect with support on the 
intersection of the M5 with the M2 along the surface $\Sigma$. The M5 world-volume theory gives rise to a Gaiotto-like field theory living on the 4-manifold $X$,
and the M2 instanton is then realized as a topological configurations of the corresponding 4d degrees of freedom having support on the surface $\Sigma\subset X$. Along  $X$, away from $\Sigma$ we have the same local physics as in absence of the M2: the $U(1)$ is Higgsed.
We claim that the 4d $U(1)$ gauge symmetry is restored along $\Sigma$. Indeed, the intersection with the M2 has a description as a topological defect in terms of the degrees of freedom living on $X$, and conversely all topologically non-trivial configurations of the 4d fields should have a geometric  engineering in terms of branes. The 4d theory is a supersymmetric version of the Abelian Higgs model;
in the broken $U(1)$ phase its topological defects are the well-known superconducting vortices \cite{WittenHiggs}. Let $z$ be a local complex coordinate so that the surface 
 $\Sigma$ is locally given by $z=0$. As $z\to0$ the gauge field behaves, in a holomorphic gauge, as $dz/z$ and the Higgs field goes to zero. Therefore the $U(1)$ 
gauge symmetry is restored along $\Sigma$. This is most easily understood from the geometric picture: the M2 brane has the effect of making the two complex planes $\{x=0\}$ and $\{y=0\}$ 
into  planes  \emph{punctured} at the respective origins, so that a new compact 1-cycle emerges, namely  the difference of the cycles around the two punctures in the respective planes.  
From the viewpoint of the field theory on the 4-manifold $X$ the gauge field associated to this 1-cycle appears to be the same one present in the original unbroken phase, before the deformation.
The compact cycle has support at the M2 brane.  Moreover
since the boundary of M2 brane sources the B-field on the M5 brane, we have
\be
dH=F\wedge  d\eta =2\pi\,\delta_{[\Sigma]}
\ee
on the M5 brane which implies $F=2\pi\,\delta_{[\Sigma]}\big|_X$.  
In other words the first equation in \eqref{kasqwer} now follows: $e(\Sigma)=c_1(\cl)$.

\subsubsection{Application to ${\cal{S}}[A_1]$ Theory}
Now we are ready to apply this setup to the Gaiotto theories. We will mainly focus on the $N=2$ case, but the generalization to all $N$ is straightforward.
As we already discussed there is a locus where the curve $C$ factorizes into two pieces.  This can be done after we adjust some of the masses of the gauge theory appropriately.
Moreover, in this limit we have $U(1)^{n-1}$ gauge factors with $n$ nodal points, i.e.\!  with $n$ charged fields where $[n-2] $ is the divisor of the one form $\lambda$ where the curve $C$ is given by $y^2= \lambda^2$.   This theory is different from the Seiberg-Witten case discussed:
We have one extra matter field compared to the number of $U(1)$'s and we can have a Higgs branch without even turning on the FI term.  However, this means that the topological theory, as the masses are tuned to allow factorization will leads to the moduli space of Seiberg-Witten equations which is not compact and which would lead to divergencies.  To avoid this divergence we can gauge an extra $U(1)$ flavor symmetry, which is available only when the masses satisfy $\sum_i\epsilon_i m_i=0$, as already discussed.  Once we gauge this $U(1)$ we will have a situation very similar to the Seiberg-Witten case discussed, namely now we have $n$, $U(1)$ gauge factors and $n$ massless charged fields, one for each nodal point.
So in this context we can repeat the exact analysis we did above for the Seiberg-Witten case, show that topological amplitudes for this weakly gauged Gaiotto theory can be captured as in the Seiberg-Witten case by pseudo-holomorphic vortices.  Roughly speaking, we get $Z_{X}=Z_{SW}^{n}$, although we expect the actual relation to  involve also contributions from contact terms and finite counterterms. 

It would be interesting to interpret the resulting amplitude for the topological theory considered here, in which one gauges an extra $U(1)$ flavor symmetry, in terms of the topological invariants computed in the original TFT without this  gauging.
As we discussed, at the point in mass parameters  where $\sum \epsilon_i m_i = 0$ the original theory is divergent (as the Seiberg-Witten equations will have non-compact moduli of solutions) and develops an extra $U(1)$ global symmetry which we gauged.  It is natural to expect that the result of the gauging the extra $U(1)$ is related to the residue of the pole in topological amplitudes in the original theory as $\sum \epsilon_i m_i \rightarrow 0$.
This may have a simple interpretation in the set-up
\cite{gukov2} where one takes the viewpoint of the 2d topological field theory living on the curve $C$, and defines the invariants of the 4-manifold in terms of correlation functions for this 2d theory. The singularity as 
$\sum \epsilon_i m_i \rightarrow 0$ gets reinterpreted as due to the collision $\lim_{z_1\to z_2}\phi_1(z_1)\phi_2(z_2)$ inside the relevant correlation. The residue is obtained by replacing the product $\phi_1(z_1)\phi_2(z_2)$ with the operator $\phi$ in the singular part of the OPE: it is a 4d topological invariant, as it is the finite part of the correlation with the pole subtracted.
The operator $\phi$ looks to have the general structure one would expect to arise in this context when we gauge the extra $U(1)$ to get our modified topological field theory. 
It would be worthwhile developing this further.

\subsubsection{Extension to general $N$}

The situation when the underlying 4d $\cn=2$ QFT is
of class $\cs[A_{N-1}]$ is similar.
At a point $u_\circ\in \cu_\circ$ where the SW curve
decomposes into $N$ linear curves,
$\prod_{\ell=1}^N(y-\lambda_\ell)$ we get a set of massless 
hypermultiplets in one-to-one correspondence with the zeros of
\be
\lambda_\ell-\lambda_{\ell^\prime}=0\qquad 1\leq \ell <\ell^\prime\leq N.
\ee
Let $q_{\ell,\ell^\prime}$ be such a zero.
One considers the M2 branes with supports of the form
\be
A_{\ell,\ell^\prime}\times \{q_{\ell,\ell^\prime}\}\subset W\mspace{-2mu}X\times T^*\mspace{-1mu}C,
\ee
where $A_{\ell,\ell^\prime}\subset W\mspace{-2mu}X$ is
an associative submanifold with boundaries on $X_{\omega_\ell}\cup X_{\omega_{\ell^\prime}}$. At a generic point in $\cu_\circ$ a Higgs branch of quaternionic dimension $N-1$ opens up, and we introduce $N-1$ abelian vector-multiplets; the corresponding $N-1$ topological $D$-terms that can be added to the action describe the independent deformations
deformations $X\to X_{\omega_\ell}$ ($\sum_\ell\omega_\ell=0$) of the co-associative supports of the branes.
At a generic point in $\cu_\circ$ we again get several copies of the Seiberg-Witten-(Taubes) equations.

\subsection{The near-symplectic case}\label{nonsym}

For simplicity we focus on the $N=2$ case.
If our 4-manifold $X$ with $b_2^+(X)\ge1$ does not admit a symplectic form, we may deform the metric so that there is a self-dual harmonic form $\Omega$ whose zero-locus $Z=\{\Omega=0\}\subset X$  is a single embedded circle  $S^1$ \cite{taubeszero}.
Then the two M5 branes $L_{t\Omega}$ and $L_-$ intersect in a collection of non-intersecting circles $S^1$ one for each
zero of the differential $\lambda$.
More generally, for a generic metric, the intersection $L_\omega\cap L_-$ consists of a set of non-intersecting embedded circles, each circle being localized at a distinct zero of $\lambda$. 

The story is similar to the symplectic case. 
The new ingredient of the analysis is the existence of additional TFT instantons of a different kind. They are given by M2 branes whose support has the form
$A\times \{\text{(a zero of $\lambda$)}\}$ with $A\subset W\mspace{-2mu}X$ an associative submanifold such that  
$A\cap X_\omega\equiv \Sigma_\omega$ and
$A\cap X\equiv \Sigma$ are pseudo-holomorphic curves with boundary on the zero set of $\omega$ 
\be
\partial\Sigma = Z= -\partial \Sigma_\omega,
\ee
and finite volume
\be
\mathrm{vol}(\Sigma)=\int_\Sigma \Omega <\infty,
\ee
a condition which may be shown to be equivalent to having finite action 
in the sense of the effective Seiberg-Witten theory \cite{taubes3}.
 
Since the co-associative deformations are trivial at the level of the 4d TFT, the computation of the usual Donaldson/Seiberg-Witten invariants should also localize around the  TFT instantons of this kind.
In particular, if the Seiberg-Witten invariant of the 4-manifold $X$ is not zero, one expects the presence of non-trivial TFT instantons of the above form. That they indeed exist is a theorem by Taubes \cite{taubes3} (see also a related theorem by Gerig \cite{gerig2}).

In fact, one expects that the full Seiberg-Witten invariants are reproduced in this way by an appropriate count of the various TFT instantons for the near-symplectic case as was the case for symplectic manifolds.
The correct count is discussed from the mathematical viewpoint in ref.\!\cite{gerig2},
where agreement is checked mod 2, but it is expected to work even by dropping the mod 2 condition.\footnote{To be clear, in ref.\!\cite{gerig1} the relevant count of (punctured) pseudo-holomorphic curves is defined over the integers, and in ref.\!\cite{gerig2} it is shown that there is a correspondence between the relevant moduli spaces of such curves and Seiberg-Witten solutions.} From the physics side it is also clear that it should work, and it would be interesting to flesh out the details of the physics that is involved in this counting.  In particular the relevant counting of the curves should be as counting embedded objects in $X$ (with multiplicities) and not as maps into $X$.  In other words, we expect the relevant invariants are the analogs of the Gopakumar-Vafa invariants rather than the Gromov-Witten invariants.

\section{Discussion} 
In this paper we have shown that embedding the ${\cal N}=2$ topological field theory on 4-manifolds into M-theory can be helpful in shedding light on the connection of Taubes' work and the Seiberg-Witten invariants, as inquired by Taubes at the end of ref.\cite{TaubesGrSW}. In particular we find that $G_2$ geometry on the space of self-dual 2-forms over the 4-manifold $X$ is necessary for this realization.  The M2 branes suspended between M5 branes realizes the Taubes' realization of Seiberg-Witten invariants as Gromov invariants for symplectic manifolds.
This setup naturally generalizes to the case of near-symplectic manifolds, where M-theory ingredients guarantee that there should be an extension for this picture, which mirrors what has been found mathematically.  Namely one ends up considering M2 branes which project to Riemann surfaces ending on zero loci of self-dual 2-forms. It would be interesting to further study the physics of this theory, as it involves superconducting vortices ending on defects.

\section*{Acknowledgments} 
   We thank Edward Witten for drawing our attention to this question, and to Sergei Gukov, Marcos Mari\~no and Pavel Putrov for helpful discussions and clarifications.
   
   The research of CV is supported in part by the NSF grant PHY-1719924 and by a grant from the Simons Foundation (602883, CV). The research of CG is supported by the National Science Foundation under Award \#1803136.

\appendix{}
\section{Appendix: Review of Taubes' results ($\omega$ symplectic)}\label{revtau}

Seiberg-Witten monopole equations \cite{Witten:1994cg} describe the supersymmetric configurations of topological twisted $\cn=2$ SQED with one massless charged hypermultiplet which can be seen as the IR effective theory of $\cn=2$ pure SYM at a point in the Coulomb branch where the monopole (or the dyon) get massless \cite{Seiberg:1994rs}.\footnote{\ 
We stress that while $\cn=2$ SQED satisfies conditions C1, C2 of \S.\ref{setup} (it is the unique theory with $C=\mathbb{P}^1$ and a single pole with $n_1=3$), $\cn=2$ pure SYM with gauge group $SU(2)$ does not satisfy them (since $\phi_2$ has pole of odd order)
so pure SYM does not admit our co-associative deformation
and its story is rather different.}

In the notation of ref.\!\cite{Witten:1994cg}
the Seiberg-Witten (SW) equations read
\be\label{sweq}
\Dsl M=0,\qquad F^+_{\mu\nu}+\frac{i}{2} M\Gamma_{\mu\nu}\bar M=0,
\ee
where the ``monopole'' $M$ is a section 
of the positive chirality sub-bundle $S_+$ of a $\mathrm{Spin}^\C$-structure on the smooth oriented 4-manifold $X$, $\Dsl$ is the Dirac operator for a connection  on $S_+$, and  $F^+$ is the self-dual part of the curvature of the induced $U(1)$ connection $A$ on $\det(S_+)$. The SW invariant
$\mathsf{SW}$ associates to each choice of the
$\mathrm{Spin}^\C$-structure an integer which ``counts'' with signs the solutions to the SW equations \eqref{sweq}. 
\smallskip

The second equation in \eqref{sweq} is just
$\delta \chi=0$ with the auxiliary field $D$ replaced by its on-shell expression using its equations of motion.
Therefore if we add to the action a topological FI term \eqref{oop0} the second SW equation gets shifted by $\omega=t\Omega$ 
 \be\label{ddefeeq}
 \Dsl M=F^++\frac{i}{2} M\Gamma\bar M+t\,\Omega=0
 \ee
 a deformation of the SW equations already considered in Witten's original paper \cite{Witten:1994cg} in order to get a better behaved one (for $t\neq0$ the gauge group acts freely on the space of solutions). 

 \medskip

 Taubes considers the deformed SW equations \eqref{ddefeeq} 
 when $X$ is a symplectic 4-fold \cite{tau1,tau2,tau3}; his analysis is nicely summarized in \S3 of ref.\!\cite{WittenHiggs}.  
When $X$ is symplectic, we may write\footnote{\ Here $K\equiv T^{(2,0)}$ is the canonical bundle as in \S\ref{se:sym}.} 
\be
S_+=E\oplus (K^{-1}\otimes E)
\ee 
for some line bundle $E$;
  the $\mathrm{Spin}^\C$-structure is specified by the Chern class $e\equiv c_1(E)$. Then the SW invariant may be seen as a map 
  \be
  \mathsf{SW}\colon H^2(X,\Z)\to\Z,\qquad e\mapsto \mathsf{SW}(e).
  \ee
The monopole field $M$ takes the form
\be
M=\begin{pmatrix} \alpha\\ \beta\end{pmatrix},
\qquad \text{where }\ \alpha\in C^\infty(E),\quad \beta\in \Lambda^{(0,2)}(E).
\ee 
Taubes studies the behavior of the solutions for $t\ggg0$.
He finds \cite{tau2,tau3}:
\begin{itemize} 
\item[a)] for  $[\Omega]\cdot e<0$
 there is no solution, so $\mathsf{SW}(e)=0$; 
 \item[b)] if $[\Omega]\cdot e=0$
 the only solution is the trivial one: zero gauge field, $\beta=0$ and $|\alpha|$ is the constant such that $M\Gamma\bar M+2i t\Omega=0$. Then $\mathsf{SW}(e)=1$;
 \item[c)] if $[\Omega]\cdot e \geq [\Omega]\cdot c$ we reduce to the above two cases by the ``charge conjugation'' symmetry 
 \be\label{chargeconj}
 \mathsf{SW}(e)=\pm\mathsf{SW}(c-e);
 \ee
 \item[d)] for $e$ in the window $0 < [\Omega]\cdot e <[\Omega]\cdot c$ we may have non-trivial solutions.
 \end{itemize}
 The interesting solutions have the following form
 (see also \cite{WittenHiggs}):
as $t$ gets large and positive, $\beta\to 0$ everywhere while $|\alpha|$ goes ``almost everywhere'' to the constant in b);  but $\alpha\in C^\infty(E)$ is forced by topology to have a non-trivial zero locus $\Sigma\subset X$
 which (by definition) is Poincar\'e dual to the Chern class $e$ of $E$.
In the limit $t\to\infty$ the zero-locus $\Sigma$ approaches a 
pseudo-holomorphic curve; indeed for $\beta=0$ the first equation \eqref{sweq} reduces to
\be
\overline{\partial}_A \alpha=0
\ee
where $\overline{\partial}_A$ is the (0,1)-part of the 
covariant derivative on $C^\infty(E)$.
It follows that in the symplectic case counting solutions to the SW equations for the $\mathrm{Spin}^\C$ structure $e$ is the same as counting pseudo-holomorphic curves $\Sigma$ in the homology class dual to $e$. In other words, in the symplectic case with $b_2^+(X)>1$ the SW invariant $\mathsf{SW}$ coincides with the Gromov invariant \cite{tau1,tau2,tau3}. 
The action of a solution to the equation \eqref{ddefeeq} for $t\gg0$ is proportional to
\be\label{jzxxxv}
t\int \Omega\wedge F/2\pi = t\,\mathrm{vol}(\Sigma),
\ee 
as follows from \S\ref{fififi}.

\begin{rem}
We clarify why it is convenient to assume $b_2^+(X)>1$.
To have a well-defined invariant, there should not be reducible solutions to the SW equations, i.e.\! solutions with $M=0$. If $b_2^+(X)\geq 1$ then there are no reducible solutions for a generic deformation $t\,\Omega$. If $b_2^+(X)>1$ there are no reducible solutions along a generic one-parameter family of deformations, so that we may reach the limit $t\to\pm\infty$ where the analysis simplifies without crossing troublesome points.
\end{rem}


\end{document}